\documentclass[twocolumn,american,prb,final,aps,superscriptaddress]{revtex4-1}
\usepackage[T1]{fontenc}
\usepackage[latin9]{inputenc}
\usepackage{xcolor}
\usepackage{array}
\usepackage{multirow}
\usepackage{amsmath}
\usepackage{graphicx}
\PassOptionsToPackage{normalem}{ulem}
\usepackage{ulem}

\makeatletter

\providecommand{\tabularnewline}{\\}
\providecolor{lyxadded}{rgb}{0,0,1}
\providecolor{lyxdeleted}{rgb}{1,0,0}

\DeclareRobustCommand{\lyxsout}[1]{\ifx\\#1\else\sout{#1}\fi}

\usepackage{pslatex}
\allowdisplaybreaks

\usepackage{babel}

\makeatother

\usepackage{babel}
\begin{document}
\title{Charge effects in donor doped perovskite ferroelectrics}
\author{J. Liu}
\affiliation{State Key Laboratory for Mechanical Behavior of Materials \& School
of Microelectronics, Xi'an Jiaotong University, Xi'an 710049, China}
\author{L. Liu}
\affiliation{College of Materials Science and Engineering, Guilin University of
Technology, Guilin, 541004, China}
\author{J. Zhang}
\affiliation{School of Microelectronics \& State Key Laboratory for Mechanical
Behavior of Materials, Xi'an Jiaotong University, Xi'an 710049, China}
\author{L. Jin}
\affiliation{Electronic Materials Research Laboratory, Key Laboratory of the Ministry
of Education \& International Center for Dielectric Research, School
of Electronic and Information Engineering, Xi'an Jiaotong University,
Xi'an 710049, China}
\author{D. Wang}
\email{dawei.wang@xjtu.edu.cn}

\affiliation{School of Microelectronics \& State Key Laboratory for Mechanical
Behavior of Materials, Xi'an Jiaotong University, Xi'an 710049, China}
\author{J. Wei}
\affiliation{Electronic Materials Research Laboratory, Key Laboratory of the Ministry
of Education \& International Center for Dielectric Research, School
of Electronic and Information Engineering, Xi'an Jiaotong University,
Xi'an 710049, China}
\author{Z.-G. Ye}
\affiliation{Electronic Materials Research Laboratory, Key Laboratory of the Ministry
of Education \& International Center for Dielectric Research, School
of Electronic and Information Engineering, Xi'an Jiaotong University,
Xi'an 710049, China}
\affiliation{Department of Chemistry and 4D LABS, Simon Fraser University, Burnaby,
British Columbia, V5A 1A6, Canada}
\author{C.-L. Jia}
\affiliation{School of Microelectronics \& State Key Laboratory for Mechanical
Behavior of Materials, Xi'an Jiaotong University, Xi'an 710049, China}
\affiliation{\textsuperscript{}Peter Grünberg Institute and Ernst Ruska Center
for Microscopy and Spectroscopy with Electrons, Research Center Jülich,
D-52425 Jülich, Germany}
\date{\today}
\begin{abstract}
Doping is a widely used method to tune physical properties of ferroelectric
perovskites. Since doping can induce charges due to the substitution
of certain elements, charge effects shall be considered in doped samples.
To understand how charges can affect the system, we incorporate the
dipole-charge interaction into our simulations, where the pinched
hysteresis loops can well be reproduced. Two charge compensation models
are proposed and numerically investigated to understand how lanthanum
doping affect BaTiO$_{3}$'s ferroelectric phase transition temperature
and hysteresis loop. The consequences of the two charge compensation
models are compared and discussed.
\end{abstract}
\maketitle

\section{Introduction}

Barium titanate (BaTiO$_{3}$) is a prototype perovskite type crystal
with good ferroelectric and dielectric properties. It is widely used
in a variety of devices such as piezoelectric and ultrasonic actuators,
pyroelectric detectors, posistors, multilayer ceramic capacitors,
temperature sensors and controllers, as well as tunable elements in
microwave circuits \citep{Choi,Kasap}. Doping BaTiO$_{3}$ with other
chemical elements is an important way to further modifying their properties
or improving their performances \citep{Buscaglia,Chen}. As a matter
of fact, trivalent rare earth (RE) elements are commonly used as dopants
for ferroelectric perovskite BaTiO$_{3}$\citep{Ihrig,Ihrig_valence}.

Depending on the valence state of the doped elements, doping can be
divided into two categories: (i) Acceptor doping where the dopants
own less ionic charge than the ion that they replace, and (ii) Donor
doping where the dopants own more ionic charge. For the acceptor doping
case, the charge balance is usually kept by oxygen vacancies, denoted
by $V_{O}^{\cdot\cdot}$. These elements usually include large, monovalent
ions substituting the A-sites, or small trivalent or divalent ions
substituting the B-sties\citep{Freeman_RE,Eichel,Jonker}. For the
donor doping case, large trivalent cations substitutes some of the
original A-site ions, or small pentavalent or hexavalent cations substituting
some B-site ions. The charge balance is usually kept by A-vacancies,
B-vacancies, or elements valance change.

Considering the intermediate size and charge of the RE$^{3+}$ ions
comparing to the Ba$^{2+}$ and Ti$^{4+}$ ions, several aliovalent
doping mechanisms (ionic and electronic) are possible. In general,
based on the size of the ions, RE ions could enter A-site (substituting
Ba) or B-site (substituting Ti) in BaTiO$_{3}$: (i) Small RE$^{3+}$
ions such as Yb$^{3+}$ can exclusively enter the B-site with charge
compensation by oxygen vacancies\citep{Buscaglia_La_TE}; (ii) Intermediate
size RE$^{3+}$ ions, such as Dy, Ho, Er, can enter the A-site or
B-site, forming systems like $\left(\textrm{Ba}_{1-x},\textrm{Re}_{x}\right)\left(\textrm{Ti}_{1-y},\textrm{Re}_{y}\right)\textrm{O}{}_{3}$,
with charge balance fulfilled by the so-called self-compensation mechanism\citep{Buscaglia,Freeman_RE,Tsur};
(iii) Large RE$^{3+}$ ions such as La, Sm dope exclusively on the
A-site\citep{Paunovi=000107,Freeman}. To make things even more complicated,
the doping mechanism is also affected by the experimental process,
such as the oxygen partial pressure ($P_{\textrm{O}_{2}}$)\citep{Albertsen},
sintering temperature/time and the overall A/B ratio in the raw materials
before sintering\citep{Eichel,Jonker,Lee_Ba/Ti}.

Doping elements of nonequivalent valence state to a ferroelectric
perovskite will inevitably induce effective charges in the system.
It is therefore important to understand how the additional charges
can affect the system, in particular their interaction with dipoles,
and the subsequent influence on the phase transitions. In this work,
we focus on the La-donor doping of BaTiO$_{3}$, which has attracted
many theoretical and experimental studies \citep{Ihrig_valence,Freeman_RE,Lewis}.
However, it is still one of the least understood aspects of the defect
chemistry in solid state compounds\citep{Freeman,Lewis,Makovec,Makovec-1}.
One important question is to determine the primary charge compensation
mechanism. Due to its larger ion size, La ions (1.36 Å) can only occupy
the A-sites (i.e., substituting Ba atoms)in the BaTiO$_{3}$ lattice.
On the other hand, the charge compensation mechanism is still a controversial
problem. When La$^{3+}$ replaced Ba$^{2+}$ on the A-site, the system
must be compensated by either cation vacancies (A- or B-sites), free
electrons, or the changed of valence state of Ti ions (i.e., Ti$^{4+}$
to Ti$^{3+}$). Since the way of charge compensation would influence
the raw material ratio and physical properties of the resulting materials
\citep{Lee_Ba/Ti,Zubko}, understanding the compensation mechanism
can help the material preparation.

With the previous investigation on the positive temperature coefficient
of resistance(PTCR) of La-doped BTO, it is widely accepted that ionic
compensation ($4\textrm{Ba}_{\textrm{Ba}}^{\times}+\textrm{Ti}^{4+}\Rightarrow4\textrm{La}_{\textrm{Ba}}^{3+}+V_{\textrm{Ti}}^{''''}$)
is the primary mechanism for high dopant concentration\citep{Morrison}.
Meanwhile, lattice energy calculations and the ternary phase diagram
of BaO-TiO$_{2}$-LaO$_{1.5}$ also support the Ti-vacancy compensation
mechanism. \citep{Freeman,Lewis,Morrison}. On the other hand, experimental
studies indicate that the electronic compensation mechanism ($\textrm{Ba}_{\textrm{Ba}}^{\times}\Rightarrow\textrm{La}_{\textrm{Ba}}^{\bullet}+e'$)
are also plausible and, in fact, preferred \citep{Ianculescu}(Ref:
Wei Paper). As these two mechanism corresponds to two stoichiometry
formula, both of which are valid, it seems hard to resolve this problem
solely by experiments. Here, we propose two models to investigate
the properties of La-doped BTO and determine if one compensation mechanism
is more likely than the other. The goal of this work is providing
us a comprehensive understanding charges and macroscopic ferroelectric
properties under doping condition on the atomic scale.

This paper is organized as the follows. In Sec. \ref{sec:Eq_and_Method},
we introduce the charge-dipole interaction to effective Hamiltonian
method and our theoretical models used for numerical simulation. Meanwhile,
we well reproduced pinched hysteresis loop of acceptor doping. In
Sec. \ref{sec:Results} and Sec. \ref{sec:discussion}, we apply Monte
Carlo (MC) calculation to BaTiO$_{3}$ and numerically obtain the
results of dopant charges. A detailed comparison is discussed. Finally,
in Sec. \ref{sec:Conclusion}, we present a brief conclusion.

\section{model and method\label{sec:Eq_and_Method}}

In this section, we will build two models for simulations, corresponding
to the two possible charge compensation mechanisms. The charge effects
of the two models are reflected in the new terms added to the effective
Hamiltonian terms used to simulation the phase transitions of BaTiO$_{3}$
\citep{JLiu_JCP}.

\subsection{Models for dopant charges}

In previous investigations, charge effects have been incorporated
to understand heterovalent relaxors such as PbMg$_{1/3}$Nb$_{2/3}$O$_{3}$
\citep{Al-Barakaty}. Here, for doped BaTiO$_{3}$, we consider the
long-range dipole-dipole and charge-dipole interactions, which can
be conveniently computed using the Coulomb interaction matrix obtained
using the Ewald method \citep{Wang2019}.

As we have mentioned in the introduction, different charge compensation
mechanism had been proposed for La-doped BaTiO$_{3}$\citep{Freeman_RE,Buscaglia_La_TE,Morrison,Chan_2,PengJC,Dawson}.
It should be noted that, due to the large size of La$^{3+}$ ($1.36$\,\AA),
La$^{3+}$ can only substitute the A-sites Ba$^{2+}$ ions. Meanwhile,
there is no direct experimental evidence of the Ba-vacancy and/or
Ti-vacancy as the charge compensation mechanism \citep{Freeman}.
Given such considerations, we will focus on two most likely charge
compensation mechanisms: (i) the electron compensation ($\textrm{Ba}_{\textrm{Ba}}^{\times}\Rightarrow\textrm{La}_{\textrm{Ba}}^{\bullet}+e'$);
(ii) Ti vacancy compensation ($4\textrm{Ba}_{\textrm{Ba}}^{\times}+\textrm{Ti}^{4+}\Rightarrow4\textrm{La}_{\textrm{Ba}}^{3+}+V_{\textrm{Ti}}^{''''}$)
\citep{Morrison}, which will be called model A and model B hereafter.

\subsubsection{Dopant arrangement \label{subsec:Dopant-arrangement}}

In model A, the charge neutrality is maintained by electron compensation.
This model are preferred by experiments in which the Ti concentration
is believed not to be modified \citep{Paunovi=000107,Ianculescu,Chan_2,Ganguly},
i.e., the ratio of $\left(N_{\textrm{La}}+N_{\textrm{Ba}}\right)/N_{\textrm{Ti}}$
is kept to 1. It can be represented as:

\begin{equation}
\textrm{La}_{2}\textrm{O}_{3}+2\textrm{Ti}\textrm{O}_{2}\Rightarrow2\textrm{La}_{\textrm{Ba}}^{\bullet}+2\textrm{Ti}_{\textrm{Ti}}^{\times}+6\textrm{O}_{\textrm{O}}^{\times}+\frac{1}{2}\textrm{O}_{2}+2e'\label{eq:model-A}
\end{equation}
and leads to the general formula $\left(\textrm{Ba}_{1-x}\textrm{La}_{x}\right)$TiO$_{3}$\citep{Morrison-1}.
The above formula shows that, in the sintering process, La$_{2}$O$_{3}$
and TiO$_{2}$ react, producing La$^{3+}$, which substitutes Ba and
owns a positive charge. Production of Ti$^{4+}$ ($\textrm{Ti}_{\textrm{Ti}}^{\times}$)
and O$^{2-}$ ($\textrm{O}_{\textrm{O}}^{\times}$) entered their
usual positions without charge anomaly. Meanwhile, the process releases
O$_{2}$, and due to the requirement of charge neutrality, extra negative
charge of $2e^{-}$ is required. In Eq. \eqref{eq:model-A}, $\textrm{La}{}_{\textrm{Ba}}^{\cdot}$
indicates that the La dopant substitutes a Ba$^{2+}$ with $+e$ charge
that is compensated by an electron. The electrons would be associated
primarily with Ti$^{4+}$, while final location of this electron is
uncertain. Considering the minimization of Coulomb energy, one obvious
choice for the electron is to stay in the same cell of the La dopant,
in which the electron combines with the Ti$^{4+}$ ($\textrm{Ti}^{4+}+e^{-}\Rightarrow\textrm{Ti}^{3+}$).

On the other hand, model B fulfills charge neutrality by introducing
Ti vacancies \citep{Freeman,Morrison,Morrison-1,Zulueta}, which can
be represented as:

\begin{equation}
2\textrm{La}_{2}\textrm{O}_{3}+3\textrm{Ti}\textrm{O}_{2}\Rightarrow4\textrm{La}_{\textrm{Ba}}^{\bullet}+3\textrm{Ti}_{\textrm{Ti}}^{\times}+V_{\textrm{Ti}}^{''''}+12\textrm{O}_{\textrm{O}}^{\times}.\label{eq:model_B}
\end{equation}
The above formula shows that, during the sintering, four Ba$^{2+}$
are replaced by four La$^{3+}$ ($4\textrm{La}_{\textrm{Ba}}^{\bullet}$)
while a Ti vacancy ($V_{\textrm{Ti}}^{''''}$) appears, which is needed
for charge neutrality, giving the general formula $(\textrm{Ba}_{1-x},\textrm{La}_{x})\textrm{Ti}_{1-x/4}\textrm{O}_{3}$
\citep{Morrison-1}. Again, with energy consideration, the Ti vacancy
should be close to the four La ions.

\textbf{}
\begin{figure}[h]
\centering{}\textbf{\includegraphics[width=8cm]{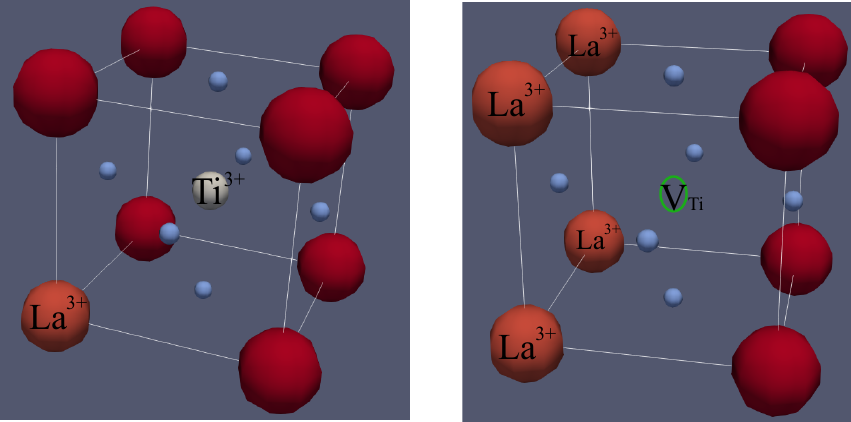}}\caption{Atom configurations for the two models: (a) Electron compensation
(model A); (b) Ti vacancy compensation (model B).\label{fig:Schematic_model}}
\end{figure}

Figure \ref{fig:Schematic_model} shows the atomic arrangements of
these two models, which is used in later MC calculations. For model
A, each $\textrm{La}_{\textrm{Ba}}^{\cdot}$ is located on the 8 corners
around a Ti$^{3+}$ with equal probability {[}see Fig. \eqref{fig:Schematic_model}(a){]};
for model B, four La$^{3+}$ ions are randomly distributed among the
8 corners of a given Ti vacancy {[}Fig. \eqref{fig:Schematic_model}(b){]},
where different choices only have minor energy differences \citep{Zulueta}.
The dipoles on the Ti vacancies is set to null ($\boldsymbol{p}_{i}=0$),
since the dipole is mostly related to Ti displacement in BaTiO$_{3}$
\citep{Zhong}. For the samples used in MC simulations, the positions
of Ti$^{3+}$ and Ti vacancies are randomly distributed.

\subsubsection{Effective charge \label{subsec:Effective-charge}}

For numerical simulation with model A and B, we also need to know
the effective charges on the La dopant and Ti ions. Such extra charge
will cause charge inhomogeneity in the system, and affect the dynamics
of dipoles, which is the focus of this work. It is important to note
that the effective charges of the relevant elements (or vacancies)
could be different from their valence states as the Born effective
charge has shown \citep{Zhong,Nishimatsu,Ghosez}. To determine the
values, we adopted a pragmatic approach by comparing simulation results
to experimental observations that the three phase transition temperatures
converge at the dopant concentration of $p\simeq10\%$\citep{Morrison,Ganguly,Kchikech}.
With a vast number of numerical calculations (not shown here), in
model A, the effective charge for $\textrm{La}{}_{\textrm{Ba}}^{\cdot}$is
set as $+2.1\thinspace\left|e\right|$, and Ti$_{\textrm{Ti}}^{'}$
is $-2.1\thinspace\left|e\right|$respectively, which is larger than
the initially expected $\pm\left|e\right|$. For model B, the effective
charge of $\textrm{La}{}_{\textrm{Ba}}^{\cdot}$ remain the same (2.1\,$\left|e\right|$)
and, in order to achieve charge neutrality, V$_{\textrm{Ti}}^{''''}$
owns a charge of -8.4\,$\left|e\right|$. We note that, according
to first principle phonon calculation described by Ref. \citep{Nishimatsu},
the Born effective charge of Ti is 7.492\,$\left|e\right|$, which
is close to the value we adopt here.

\subsection{Effective Hamiltonian}

We use a first-principles-based effective Hamiltonian approach \citep{Zhong,Nishimatsu,Wang_2014,Wang_2016}
to obtain finite temperature properties of the doped system. A new
energy item is introduced into total energy to account for the charge-dipole
interaction. The new energy expression is given by :

\begin{align}
E^{\textrm{tot}} & =E^{\textrm{self}}\left(\left\{ \boldsymbol{u}\right\} \right)+E^{\textrm{dpl}}\left(\left\{ \boldsymbol{u}\right\} \right)+E^{\textrm{short}}\left(\left\{ \boldsymbol{u}\right\} \right)\nonumber \\
 & +E^{\textrm{elas}}\left(\left\{ \eta_{l}\right\} ,\eta_{H}\right)+E^{\textrm{int}}\left(\left\{ \boldsymbol{u}\right\} ,\left\{ \eta_{l}\right\} ,\eta_{H}\right)\nonumber \\
 & +E^{\textrm{chg-dpl}}\left(\left\{ \boldsymbol{u}\right\} ,q\right)\label{eq:Etot}
\end{align}
which consists of six parts: (i) the local-mode self-energy, $E^{\textrm{self}}\left(\left\{ \boldsymbol{u}\right\} \right)$;
(ii) the long-range dipole-dipole interaction, $E^{\textrm{dpl}}\left(\left\{ \boldsymbol{u}\right\} \right)$;
(iii) the short-range interaction between soft modes, $E^{\textrm{short}}\left(\left\{ \boldsymbol{u}\right\} \right)$;
(iv) the elastic energy, $E^{\textrm{elas}}\left(\left\{ \eta_{l}\right\} \right)$;
(v) the interaction between the local modes and local strain, $E^{\textrm{int}}\left(\left\{ \boldsymbol{u}\right\} ,\left\{ \eta_{l}\right\} \right)$;
(vi) the long range charge-dipole interaction energy, $E^{\textrm{chg-dpl}}\left(\left\{ \boldsymbol{u}\right\} ,q\right)$,
where $\boldsymbol{u}$ is the local soft-mode amplitude vector (directly
proportional to the local polarization), $q$ is the charge induced
by doped unequivalent-valence ions, and $\eta_{H}$ ($\eta_{l}$)
is the six-component homogeneous (inhomogeneous) strain tensor in
Voigt notation \citep{Zhong}. The parameters appearing in the effective
Hamiltonian have been reported in Ref. {[}\onlinecite{Nishimatsu}{]}
with the effective charges being discussed in Sec. \ref{subsec:Effective-charge}.
We further assume that the charges are fixed in space, the long range
charge-charge Coulomb interaction energy, $E^{\textrm{chg-chg}}\left(q\right)$
is not included in the effective Hamiltonian. The charge-dipole energy
is treated with the Ewald method and the details can be found in Ref.
\onlinecite{Wang2018}.

With the above effective Hamiltonian, we perform MC simulations with
a pseudo-cubic supercell of size $\mathrm{12\times12\times12}$ (i.e.,
1728 unit cells) with periodic boundary conditions. Among the supercell,
the dopants and vacancies are placed according to the arrangement
described in Sec. \ref{subsec:Dopant-arrangement} and the dopant
concentration is calculated with respect to by La ions, which is from
$0\,\%$ to $10\,\%$ with an step of $1\,\%$. For each of the doped
samples, we gradually cool down the system from high temperature(\textasciitilde 550\,K)
to low(\textasciitilde 30\,K) with a step of 10\,K using 320,000
MC sweeps for each temperature to obtain the equilibrium state and
thus the local mode $\left\{ \boldsymbol{u}\right\} $, which is used
to obtain the polarization of the system at the given temperature.

\section{Results\label{sec:Results}}

Having shown the models and the effective Hamiltonian, we proceed
to use MC simulations to obtain basic properties of La donor doping
BaTiO$_{3}$, including hysteresis loops, polarization, as well as
phase transition temperatures, with different charge concentration
and configurations. Before showing such results, we firstly certify
our numerical scheme, including the Ewald interaction matrix and the
newly developed MC program with $E^{\textrm{chg-dpl}}\left(\left\{ \boldsymbol{u}\right\} ,q\right)$,
against the defect dipole model \citep{RenXB,Cohen_PRL,Cohen_APL},
which is known to gives rise to the pinched hysteresis loop.

\subsection{Pinched hysteresis loop\label{sec:Charge_effect-1}}

It is generally accepted that the oxygen-vacancy plays an important
rule in the formation of pinched and double hysteresis loops for acceptor
doping\citep{RenXB}. Ren proposed that the defect dipole is microscopically
formed by oxygen vacancies and an impurity ions within one unit cell
\citep{RenXB}. Such intentionally oriented dipoles (defect dipoles)
are employed to investigate electrocaloric effect of BaTiO$_{3}$\citep{Xv_BaiXiang}.
Cohen \emph{et al}. used oriented dipoles to investigate the aging
process of ferroelectric ceramics \citep{Cohen_PRL,Cohen_APL}, and
the key is that some dipoles are fixed to one particular direction
without fluctuation\citep{RenXB,Cohen_PRL,Cohen_APL}.

\begin{figure}[h]
\centering{}\includegraphics[width=6cm]{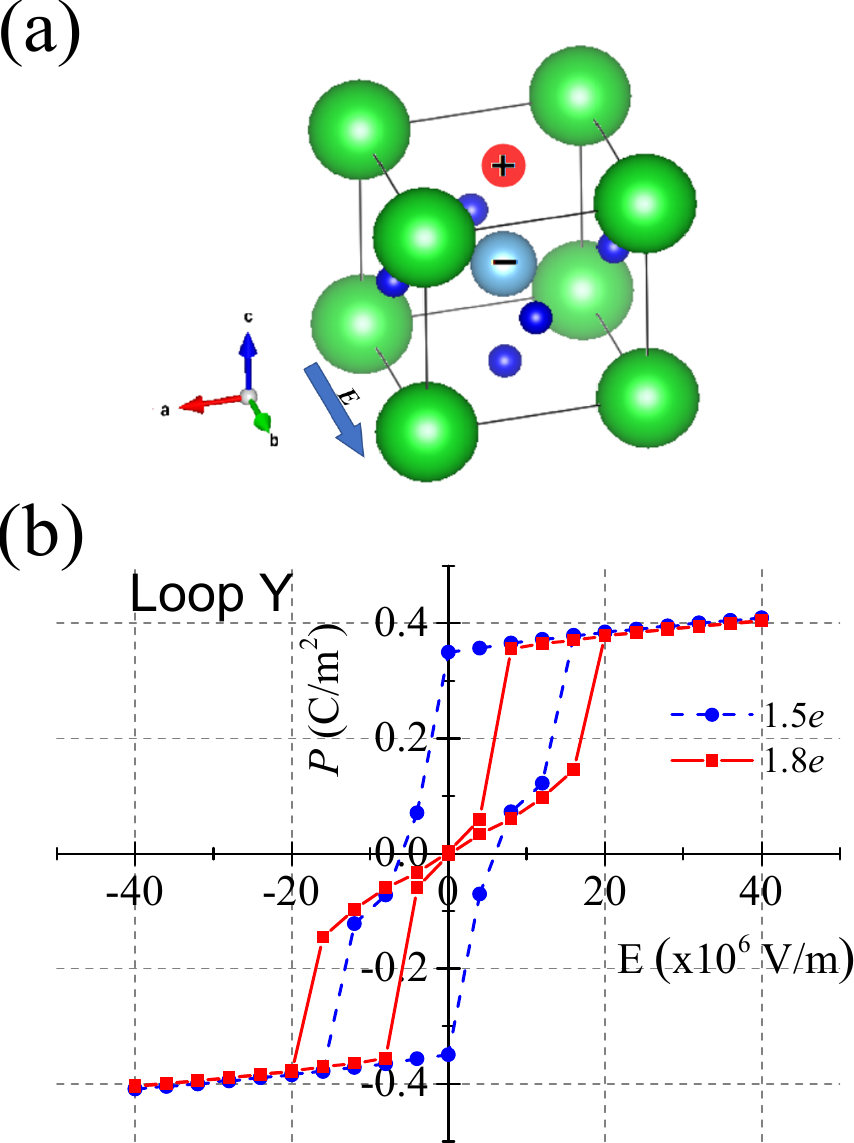}\caption{(a) Schematic drawing of defect dipole models ({[}001{]}) for acceptor
doping. (b) Simulated hysteresis loops for BaTiO$_{3}$ with effective
charge $q=1.5\,\left|e\right|$ and $q=1.8\thinspace\left|e\right|$,
at the dopant concentration $p=1.2\,\%$, $T=300$\,K. \label{figs:Cohen_pinched_loops}}
\end{figure}
 Instead of using defect dipoles, we can deal with oxygen vacancies
and impurity ions directly using the Ewald method \citep{Wang2019}.
For each unit cell with defects, we set a positive charge $+q$ on
the oxygen vacancy ($\textrm{V}_{\textrm{O}}^{\bullet\bullet}$) in
a certain direction (e.g. the {[}001{]} direction) from the Ti ion,
and a negative charge, $-q$, on the B site if it is an impurity ion
\citep{Wang2019,Zhong} {[}see Fig. \ref{figs:Cohen_pinched_loops}(a){]}.
The electric field generated by the charges on dipoles are incorporated
in the charge-dipole interaction matrix \citep{Wang2019}. The model
used here is similar to Model A, except that in model A, the positive
charge ($\textrm{La}_{\textrm{Ba}}^{\bullet}$) is randomly distributed
on the 8 corners of the unit cell.

Figure \ref{figs:Cohen_pinched_loops}(b) shows the result of the
calculated hysteresis loop at 300\,K with a sample of 1.2\,\% doping
concentration. The shape of the hysteresis loop starts to show pinched
effect at $q=1.5\thinspace\left|e\right|$, which become pronounced
when $q=1.8\thinspace\left|e\right|$, consistent with the results
of Liu \emph{et al} (see Fig. 2 in Ref. \citep{Cohen_APL}). However,
dealing with charges directly provide us more freedom in dealing with
doping effects.

It is worth noting that defect dipoles are usually assumed to be fixed
along some fixed directions (e.g., the {[}001{]} direction), making
other dipoles easier and align and the ferroelectric phase transition
temperature higher, which is inconsistent with many experimental result
\citep{Xv_BaiXiang,Nishimatsu_defects}. This challenge and its possible
solution will be discussed in more detail elsewhere.

\subsection{Energy}

\begin{table}[h]
\noindent \centering{}%
\begin{tabular}{cccc}
\hline 
Energy (Hartree) & Model A & Model B & BaTiO$_{3}$\tabularnewline
\hline 
\hline 
Self energy & 17.7758 & 19.0218 & 20.1890\tabularnewline
\hline 
Short range & -3.2441 & -3.2003 & -4.0979\tabularnewline
\hline 
Elastic & 2.8230 & 2.8547 & 3.3897\tabularnewline
\hline 
Dipole-elastic & -1.9692 & -2.0801 & -2.9728\tabularnewline
\hline 
Dipole-dipole & -9.7998 & -10.9265 & -13.6581\tabularnewline
\hline 
Charge-dipole & 0.2181 & -0.0403 & -\tabularnewline
\hline 
Total energy & 5.8038 & 5.6293 & 2.8499\tabularnewline
\hline 
\end{tabular}\caption{Constituent energy of the effective Hamiltonian for 3\% doped samples
relaxed at 300 K for model A and model B.\label{tab:Calculated-Coulomb-energy}}
\end{table}

Table \ref{tab:Calculated-Coulomb-energy} shows the constituent energies
from the effective Hamiltonian for the 3\% doped sample. These samples
are firstly relaxed at 300\,K (resulting in the tetragonal phase)
and then their energies calculated. For model A and B, the energies
are close to each other, except the obvious difference in the charge-dipole
interaction energy. Both doped samples show that the doping increases
the energy of the system comparing to the pure BaTiO$_{3}$. 

\subsection{$P_{s}$ and $P_{r}$}

\begin{figure}[h]
\centering{}\includegraphics[width=6cm]{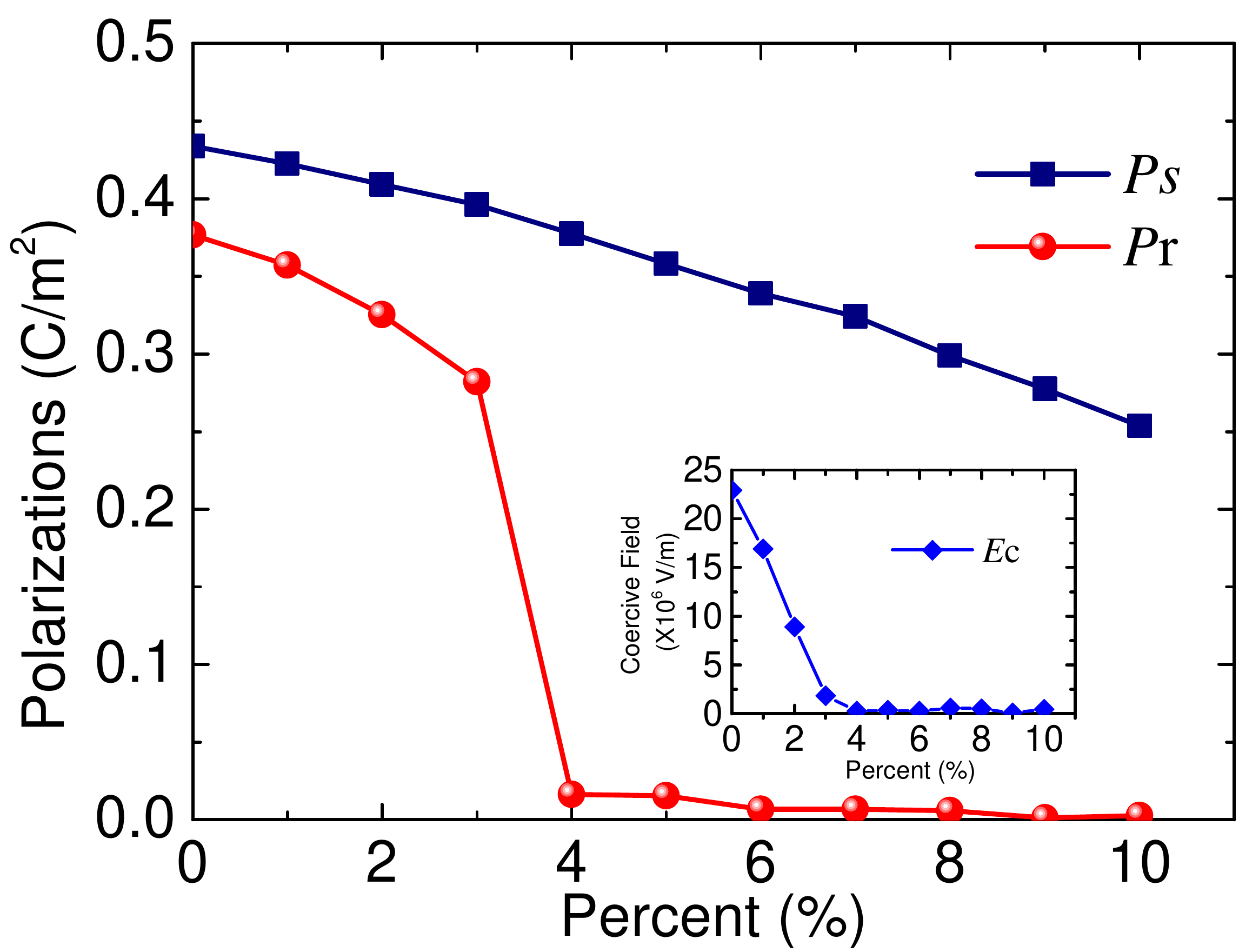}\\
\includegraphics[width=6cm]{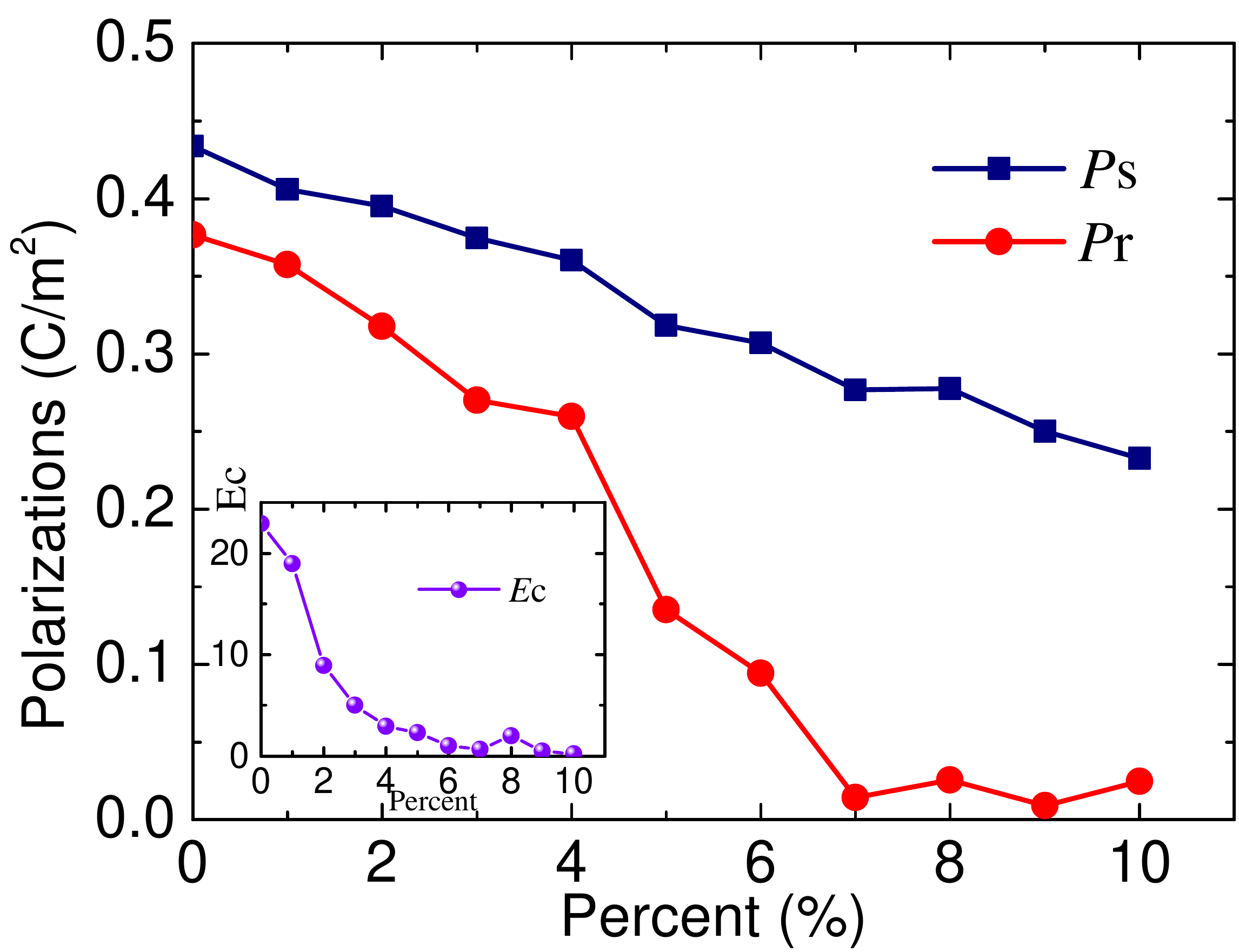}\caption{The saturation polarization (obtained at $E=5.0\times10^{7}$\,V/m),
the remnant polarization and the coercive field obtained at 300\,K
with (a) model A; (b) model B. \label{fig:PsPrEc}.}
\end{figure}

$P_{s}$ and $P_{r}$ are two important parameters to characterize
a ferroelectric material. $P_{s}$ is the saturation polarization
under a large electric field, and $P_{r}$ represents the remnant
polarization under zero electric field when the electric field is
gradually lowered. Larger $P_{s}$ usually indicates larger dielectric
constant, which are useful for capacitors, energy storage devices
and insulators \citep{Kasap,Morrison_JAP}. $P_{r}$ is often associated
with pyroelectric and piezoelectric properties \citep{Xv_BaiXiang,Nishimatsu_defects}.

Using model A and B, we have numerically calculated a series of hysteresis
loop for various doping concentrations ($0\sim10\%$) and summarized
the results in Figure \ref{fig:PsPrEc}. Pure BTO owns largest saturation
polarization ($P_{S}$), remnant polarization ($P_{r}$), coercive
field ($E_{c}$) \citep{Ihrig,Tsur,Paunovi=000107}. The hysteresis
loops of the doped systems become much narrower with lower $P_{S}$,
$P_{r}$, and $E_{c}$. It is attributed to the internal electric
field associated with the dopants that disrupt the long range dipole-dipole
interactions \citep{JLiu_JCP,Xv_BaiXiang}.

\begin{table}[h]
\begin{centering}
\includegraphics[width=8cm]{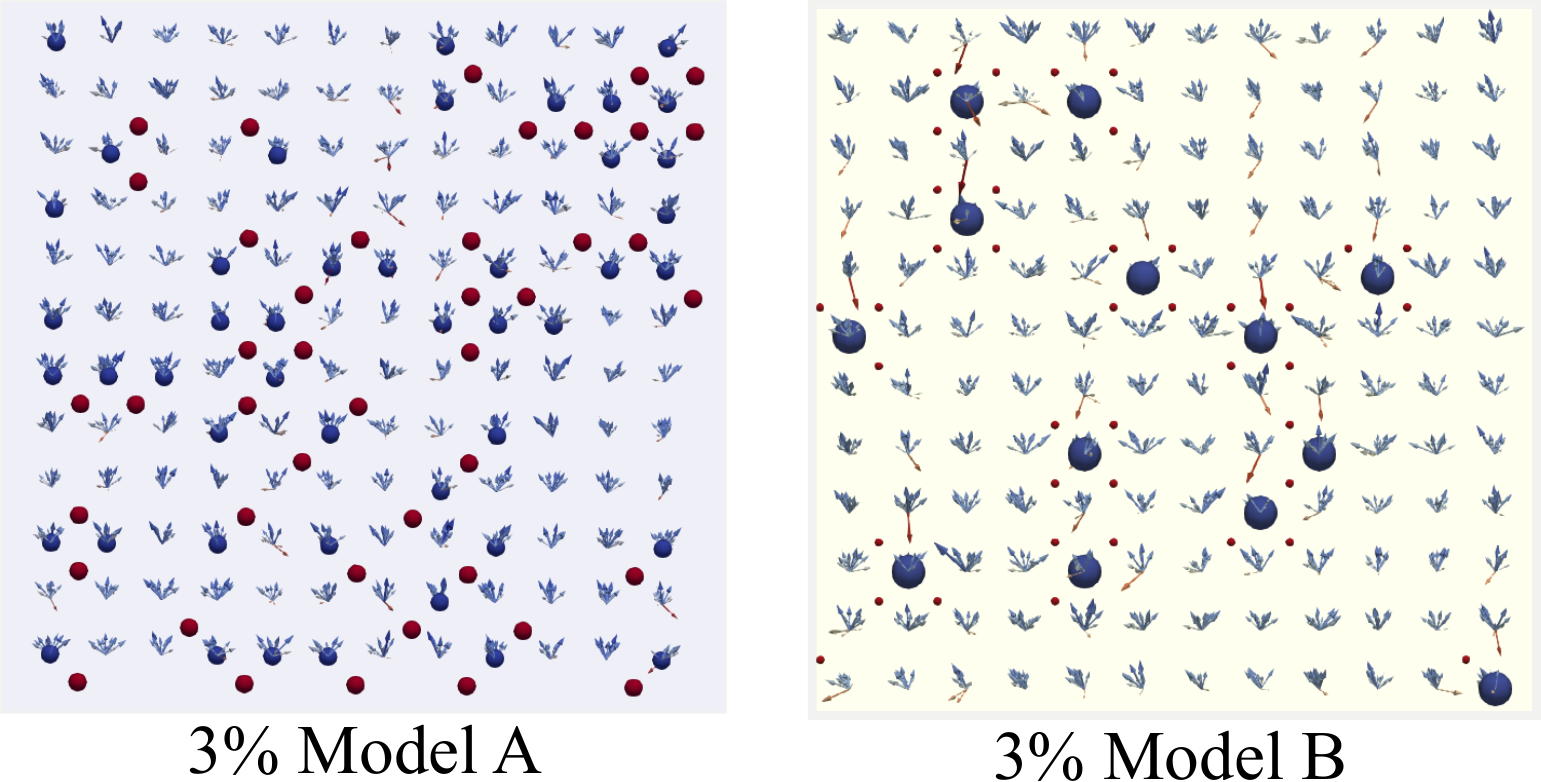}\\
\begin{tabular}{|c|c|c|c|c|}
\hline 
 &  & $N_{\textrm{AD}}$ & $\overline{\mathbf{u}_{z}}$ & $N_{V_{\textrm{Ti}}}$($\mathbf{u}_{i}=0$)\tabularnewline
\hline 
\hline 
\multirow{2}{*}{3\% 300K} & Model A & 59 & -0.011 & 0\tabularnewline
\cline{2-5} \cline{3-5} \cline{4-5} \cline{5-5} 
 & Model B & 87 & -0.021 & 13\tabularnewline
\hline 
\multirow{2}{*}{5\% 250K} & Model A & 107 & -0.014 & 0\tabularnewline
\cline{2-5} \cline{3-5} \cline{4-5} \cline{5-5} 
 & Model B & 152 & -0.023 & 22\tabularnewline
\hline 
\end{tabular}
\par\end{centering}
\caption{Dipole configuration information of doped samples.Red arrow: Active
dipole; Red ball: positive change; Blue ball: negative charge; $N_{\textrm{AD}}$:
number of active dipoles in \textit{z}-axis; $\overline{\mathbf{u}_{z}}$:
averaged \textit{z}-axis local mode value of active dipoles; $N_{V_{\textrm{Ti}}}$:
number of Ti vacancies.\label{tab:Nad_and_u}}
\end{table}
 For model A, Fig. \ref{fig:PsPrEc} (a) shows that both $P_{r}$
and $E_{c}$ have a sudden change at $p\simeq4\%$, indicating a transition
from a ferroelectric to a paraelectric state and the hysteresis loop
vanishes. For model B, Fig. \ref{fig:PsPrEc} (b) shows that $P_{r}$
and $E_{c}$ gradually decreases and vanishes at $p\simeq6\%$. We
note that the $P_{s}$ of model B {[}$P_{s}\left(B\right)${]} is
always smaller than model A for all the dopant concentration, which
can be attributed to the difference in magnitude of active dipoles
and the number of defective dipoles in the doped samples\citep{JLiu_JCP},
which is shown in Tab. \ref{tab:Nad_and_u}.

Table \ref{tab:Nad_and_u} also shows other difference between model
A and model B. At 300\,K, for the $p=3\%$ doped samples, the averaged
local mode of the induced active dipoles in model B is $\left\langle \mathbf{u}_{z}\right\rangle \simeq-0.021$
(in opposite direction to the overall polarization), which is twice
as larger as Model A. The number of active dipoles ($N_{\textrm{AD}}$)
for model A and B are 59 and 87, respectively. In addition, there
are 13 defective dipoles ($\boldsymbol{p}_{i}=0$) in model B, which
are caused by the missing of Ti ions. Therefore, two factors explain
why $P_{s}\left(B\right)$ is much smaller than $P_{s}\left(A\right)$
even for higher dopant concentrations: (i) The number and magnitude
of active dipoles in model B are larger than that in model A; (ii)
In model A, we do not assume that dipoles are missing on the dopant
sites as in model B, where the dipole is ruined when V$_{\textrm{Ti}}^{''''}$
is present while fewer dipoles usually means weaker ferroelectricity.
The observation that $P_{r}\left(A\right)$ shows a faster decrease
is likely due to the wider spread of the dopants in model A. Such
dopants distribution may effectively creates polar nanoregions that
introduces relaxor behaviors into the system.

\subsection{Ferroelectric phase transition}

\begin{figure}[h]
\centering{}\includegraphics[width=8cm]{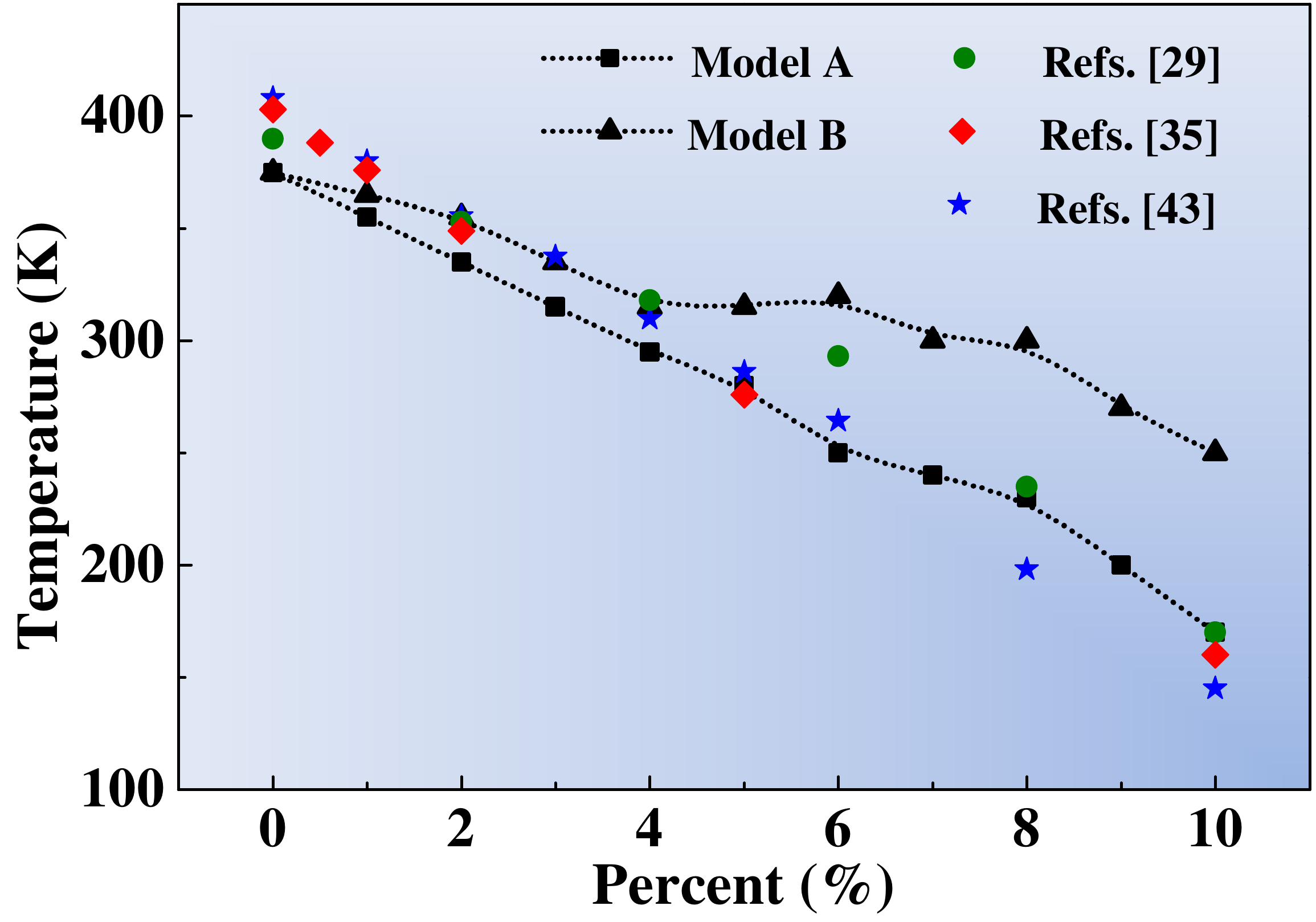}\caption{Numerically obtained ferroelectric phase transition temperature $T_{C}$
versus dopant concentration using model A and model B are shown along
with experimental values.\label{fig:Exps_vs_Cal}}
\end{figure}

Pure BTO has the ferroelectric to paraelectric phase transition at
$T=375$\,K\citep{Nishimatsu}. When La is doped into BTO, the phase
transition temperature $T_{C}$ changes with dopant concentration
$p$. Figure \ref{fig:Exps_vs_Cal} shows the how the phase transition
temperature change with dopant concentration for model A and B, and
their comparison to experimental results. The results show that for
both model A and model B $T_{C}$ decreases rapidly with the dopant
concentration. Model A has a faster decrease from $375$\,K (at $p=0\%$)
to 170\,K (at $p=10\%$), which agrees with the trend shown in experiments
\citep{Ganguly}. Model B shows the same tendency for $p<4\%$. However,
the decrease with model B becomes stagnant for $p>4\%$. We note that
model B, which use Ti vacancy as the assumption, cannot produce either
of the experimental results shown in Fig. \ref{fig:Exps_vs_Cal},
even for the experiment that claims the existence of Ti vacancies
\citep{Morrison_JAP}. Possible reasons will be discussed in Sec.
\ref{sec:discussion}

\section{Discussion\label{sec:discussion}}

Despite extensive experimental studies, charge compensation induced
effects in La doped BaTiO$_{3}$ (and more generally rare-earth elements
doped ferroelectrics) is still an open question. This lack of understanding
is in part due to the various experimental results (with similar samples)
that demonstrate significant differences. Most experiments with La
doped BaTiO$_{3}$ assumes stoichiometry formula $\left(\textrm{Ba}_{1-x}\textrm{La}_{x}\right)\textrm{Ti}\textrm{O}_{3}$
to effectively understand (or explain) properties arising from doping
\citep{Paunovi=000107,Ianculescu,Vijatovi=000107,Bi}. Ganguly\textit{
et al. }prepared a variety of compositions in order to justify A-site
vacancy is possible in La doped BaTiO$_{3}$ \citep{Ganguly}. Morrsion
\textit{et al.} considers the use of less Ti (in order to induce possible
Ti vacancies) and obtained the experimental $T_{C}$ (see Fig. \ref{fig:Exps_vs_Cal}).

Makovec \textit{et al.} prepared different samples (according to the
two different formula) by the conventional mixed-oxide method and
studied their microstructure change during their reduction/reoxidation
process\citep{Makovec}. They have shown that when (Ba$_{1-x}$La$_{x}$)(Ti$_{1-x}^{4+}$,
Ti$_{x}^{3+}$)O$_{3}$ was exposed to oxidizing environment, electron
diffraction pattern revealed that the Ti-rich phase (Ba$_{6}$Ti$_{17}$O$_{40}$)
is expelled from the solid solution, and the donor charge compensation
changed from electronic compensation to Ti vacancy compensation \citep{Jonker,Makovec}.
Lanculescua \textit{et al.} prepared the two different samples (assuming
the two different compensation mechanism) at $0.5\,\%$ dopant concentration\citep{Ianculescu}.
They showed that the phase transition temperature $T_{C}$ decrease
remarkably with all their hysteresis loops disappear, which is not
expected \citep{Ianculescu}. Meanwhile, Ti$^{3+}$ is believed to
be responsible for the dark color of the doped samples, which had
been in experiments \citep{Makovec}.

On the theory side, Lewis \textit{et al.} considered the ternary phase
diagram proposed by Jonker \textit{et al} \citep{Jonker} and the
calculated binding energy of $\left[\textrm{La}{}_{\textrm{Ba}}\textrm{V}_{\textrm{Ti}}^{''''}\right]$
\citep{Lewis}, proposed that in La doped BTO the charge compensation
is predominantly by electron, but with with high dopant concentration,
Ti vacancy appears \citep{Jonker,Lewis}. However, Freeman \textit{et
al.} keep the view that electron compensation mechanism is never the
primary one, irrespective of the donor dopant concentration \citep{Freeman,Morrison-1}.

Based on our findings, the electron transfer model (model A) is the
more likely compensation mechanism. Summarizing the available results
(including our simulation here), we have some insights regarding how
doping works. The key that dopants works is disrupting the long range
dipole-dipole interactions, which work through three ways: (i) Doping
induces extra charges in the system, producing additional internal
electric field than a pure dipole system, therefore disturbing the
dipole-dipole order of pure BTO; (ii) Precipitated compounds. Under
oxidation condition, Ba$_{2}$TiO$_{4}$ and/or Ba$_{6}$Ti$_{17}$O$_{40}$
may arise and embed in the system \citep{Lewis,Makovec,Makovec-1,Chan},
which can disrupt the long-range order as well; (iii) Local strain
that may be induced by the change of the valence state from Ti$^{4+}$
to Ti$^{3+}$ since the ionic radius of Ti$^{3+}$ is larger than
Ti$^{4+}$. However, since Ti ions are inside the oxygen octahedron,
it is hard to estimate how large the local strain shall be. While
the model here have not included all the factors, the simulation results
indicate that factor (i) is probably most important that can account
for the $T_{C}$ decrease and reproduce their properties and displays
the dipoles configurations for samples with different dopant concentration.

Our simulation results have that model A is a better model to understand
La-doped BaTiO$_{3}$, supporting the idea that electronic compensation
is the underlying mechanism. However, it should be noted that there
are limitations to our models and simulations. In model B, since Ti
vacancies exist, it may be better if the strain associated with them
is included at higher dopant concentration. This factor could be the
reason that model B cannot well reproduce the trend of $T_{C}$ for
$p>5\%$. To reveal how important such an effect can be, we further
tested a modified model B with charge and strain effects included.
However, we were not able to see a clear improvement.

\section{Conclusion \label{sec:Conclusion}}

In this work, we have considered the charge effects in ferroelectric
materilas, taking into account the long range charge-dipole interaction
\citep{Wang2019}. We first verified our approach producing the pinched
hystersis loops of acceptor doping as know from defect dipole model.
We have applied this scheme to La doped BaTiO$_{3}$ and proposed
two theoretical calculation models. With these two models, saturation
polarization, remnant polarization and coercive field are obtained,
which decrease with increasing dopant concentration. Focusing on the
paraelectric to ferorelectric phase transition temperature $T_{C}$,
we find that the disruption of long range dipolar interaction due
to the doped doponts can strongly influence the phase transition temperature.
Our results indicate that the electron compensation mechanism (model
A) is a more plausible model to understand the doping effects. We
note that our approach is also applicable to other doped situations
where the effective charge for the doped chemical element is known.
We hope this work will inspires more investigation to understand the
doping effects, which is important and interesting.
\begin{acknowledgments}
This work is financially supported by the National Natural Science
Foundation of China (NSFC), Grant No. 11574246, U1537210, and 11974268.
The authors also acknowledge the support by the HPC platform of Xi'an
Jiaotong University for providing computational resources. D.W. also
thanks the support from the Chinese Scholarship Council (201706285020).
\end{acknowledgments}

\end{document}